\documentclass[sigconf]{aamas} 
\usepackage{cleveref} \usepackage{enumitem}

\usepackage{flushend}   
\usepackage{xurl}
\usepackage{listings}
\usepackage{xcolor}
\usepackage[T1]{fontenc}
\usepackage{minted}
\usepackage{tabularx}
\usepackage[colorinlistoftodos]{todonotes}

    \setcopyright{none}
    \acmConference[NEXUS '26]%
    {Proc.\@ of the Neurosymbolic eXplainable Trustworthy Systems Workshop (NEXUS 2026)}%
    {May 25 -- 26, 2026}%
    {Paphos, Cyprus,https://nexus.telecom-paris.fr}%
    {Bakirtzis, Fridovich-Keil , Kakas, Theodorou, Li, McKinlay (eds.)}%
    \copyrightyear{2026}
    \acmYear{2026}
    \acmDOI{}
    \acmPrice{}
    \acmISBN{}
\graphicspath{{img/}}

\acmSubmissionID{<<submission id>>}

\title{Towards Neuro-symbolic Causal Rule Synthesis, Verification, and Evaluation Grounded in Legal and Safety Principles}

\author{Zainab Rehan}
\orcid{0009-0005-9336-9505}
\affiliation{
 \institution{Hasso Plattner Institute \\University of Potsdam}
 \streetaddress{Prof.-Dr.-Helmert Str. 2-3, D-14482}
 \country{Potsdam, Germany}
}
\email{zainab.rehan@uni-potsdam.de}

\author{Christian Medeiros Adriano}
\orcid{0000-0003-2588-9937}
\affiliation{
 \institution{Hasso Plattner Institute \\University of Potsdam}
 \streetaddress{Prof.-Dr.-Helmert Str. 2-3, D-14482}
  \country{Potsdam, Germany}
}
\email{christian.adriano@hpi.de}

\author{Sona Ghahremani}
\orcid{0000-0003-0697-9195}
\affiliation{
 \institution{Hasso Plattner Institute \\University of Potsdam}
 \streetaddress{Prof.-Dr.-Helmert Str. 2-3, D-14482}
  \country{Potsdam, Germany}
}
\email{sona.ghahremani@hpi.de}

\author{Holger Giese}
\authornote{IEEE and ACM Member}
\orcid{0000-0002-4723-730X}
\affiliation{
 \institution{Hasso Plattner Institute \\University of Potsdam}
 \streetaddress{Prof.-Dr.-Helmert Str. 2-3, D-14482}
  \country{Potsdam, Germany}
}
\email{holger.giese@hpi.de}

\begin{abstract}
Rule-based systems remain central in safety-critical domains but often struggle with scalability, brittleness, and goal misspecification. These limitations can lead to reward hacking and failures in formal verification, as AI systems tend to optimize for narrow objectives. In previous research, we developed a neuro-symbolic causal framework that integrates first-order logic abduction trees, structural causal models, and deep reinforcement learning within a MAPE-K loop to provide explainable adaptations under distribution shifts. In this paper, we extend that framework by introducing a meta-level layer designed to mitigate goal misspecification and support scalable rule maintenance. This layer consists of a Goal/Rule Synthesizer and a Rule Verification Engine, which iteratively refine a formal rule theory from high-level natural-language goals and principles provided by human experts. The synthesis pipeline employs large language models (LLMs) to: (1) decompose goals into candidate causes, (2) consolidate semantics to remove redundancies, (3) translate them into candidate first-order rules, and (4) compose necessary and sufficient causal sets. The verification pipeline then performs (1) syntax and schema validation, (2) logical consistency analysis, and (3) safety and invariant checks before integrating verified rules into the knowledge base. We evaluated our approach with a proof-of-concept implementation in two autonomous driving scenarios. Results indicate that, given human-specified goals and principles, the pipeline can successfully derive minimal necessary and sufficient rule sets and formalize them as logical constraints. These findings suggest that the pipeline supports incremental, modular, and traceable rule synthesis grounded in established legal and safety principles.
\end{abstract}


\newcommand{\BibTeX}{\rm B\kern-.05em{\sc i\kern-.025em b}\kern-.08em\TeX}

\usepackage{xurl}
\begin{document}


\pagestyle{fancy}
\fancyhead{}


\maketitle 


\section{Introduction}

\textbf{Context} -- Rule-based systems~\cite{liu2016rulebased} are specified and constructed from existing domain knowledge about desired and undesired (forbidden, unsafe) behavior, often as logical if–then statements. Rule-based approaches remain central in many safety- and mission-critical contexts because rules make system behavior observable at runtime and open to formal verification ~\cite{hayesroth1983rulebased,musen1989brittleness}. 

In self-adaptive and autonomic systems, the MAPE-K feedback loop~\cite{kephart2003autonomic} represents monitoring, analysis, planning, and execution as activities over a shared knowledge base (of rules and representation models). These systems often rely on runtime enforcement components that supervise primary controllers and trigger predefined mitigation actions when rules are violated~\cite{musen1989brittleness,almeida2017mapekguidelines}.

\textbf{Challenges} -- However, rule-based and expert systems are widely recognized for their brittleness and limited scalability. As new requirements emerge, even minor changes can cause cascading revisions across numerous interdependent rules. Managing and tracking these evolving rules places a significant cognitive and organizational burden on human experts, often becoming a central bottleneck and increasing the risk of unexpected failures~\cite{musen1989brittleness,greene1987automated}. Classic analyses of earlier expert systems highlight the exponential increase in maintenance effort and the well-known “knowledge acquisition bottleneck.” In practice, large rule sets are inherently difficult to extend or adapt without unintentionally disrupting existing behaviors~\cite{greene1987automated,musen1989brittleness}. These issues were prominent in large, early AI expert-system projects, which helped motivate later shifts toward methods that combine statistical learning with more scalable forms of knowledge representation~\cite{lenat1990using,musen1989brittleness}.

\textbf{Approach} -- We extend a prior neuro-symbolic causal framework for self-adaptive, learning-enabled systems\cite{adriano2024principled, adriano2025neuro} with a meta-level synthesis and verification layer that incrementally refines the governing rule theory. A Goal/Rule Synthesizer uses large language models to decompose high-level, natural-language goals and principles into candidate causes, consolidate their semantics, and translate them into candidate first-order rules while identifying necessary and sufficient cause sets. A complementary Rule Verification Engine then enforces syntax and schema correctness, logical consistency, and safety and invariant constraints before integrating only verified rules into the knowledge base. Together, these components realize an incremental, modular, and traceable pipeline from human-specified legal and safety principles to formally verified rule sets, demonstrated on autonomous driving scenarios.

\textbf{Evaluation} -- We present an application in the domain of autonomous driving, where safety goals are systematically decomposed into subgoals and translated into corresponding rules. The mapping between goals and rules is realized via simulated abduction that produces explanations of an effect from its causes, while rule refinement is performed through deductive reasoning, de-duplicating rules and combining them into necessary and sufficient sets. We evaluated our approach with a proof-of-concept implementation in two autonomous driving scenarios, showing that, given human-specified goals and principles, the pipeline can successfully derive minimal necessary and sufficient rule sets and formalize them as logical constraints. These findings suggest that the pipeline supports incremental, modular, and traceable rule synthesis grounded in established legal and safety principles.

\textbf{Contributions} -- By structuring rule maintenance as an incremental, meta-level synthesis-and-verification loop, the approach directly attacks the scalability and brittleness problems of classic rule-based systems. First, high-level goals are decomposed into modular sets of necessary and sufficient rules per goal, so extensions localize to specific goal-linked modules instead of triggering cascading revisions across a monolithic rule base. Second, semantic consolidation, de-duplication, and explicit traceability from each rule back to its originating goal and principles reduce redundancy and make large rule sets easier to understand, audit, and evolve. Third, the Rule Verification Engine systematically filters new rules through syntax, consistency, and safety checks before reintegration, preventing brittle, ad hoc patches and enabling controlled growth of the theory over time. Reproduction package is  available at~\cite{repo2026rulesynthesis}.


\section{State of the Art}\label{sec:state_of_the_art}

\textbf{Goal Decomposition with LLMs} -- 
%
 LLMs can decompose complex goals into structured subgoals for planning and assistance. Multi-step plans are generated, refined, and evaluated~\cite{aghzal2025survey}. SGA-ACR~\cite{fan2025subgoal} produces verifiable sub-goal chains for RL agents, while DELTA~\cite{liu2025delta} leverages scene graphs for efficient long-horizon task decomposition. Hierarchical LLM agents improve tractability and interpretability~\cite{hill2025generativeworldmodelstasks}, and human-centered methods~\cite{wen2024learning} learn decompositions that enhance non-expert performance on complex programming tasks.

\textbf{Abduction and Rule Synthesis} -- 
Abductive reasoning generates explanatory hypotheses and candidate symbolic rules~\cite{Aliseda2017}, providing a principled path from observations to minimal assumptions and compact rule-like hypotheses~\cite{paul1993approaches}. Statistical learning links, via inductive logic programming and statistical relational learning, allow frequent patterns or abductive explanations to become human-interpretable if–then rules~\cite{DeRaedtKersting2008,hitzler2022neural}. Contemporary neuro-symbolic approaches use abduction as structured prior knowledge to guide neural learners, improving sample efficiency. Extracted hypotheses can be refined and pruned to produce robust, generalizable adaptation rules, bridging observed outcomes and formalized reusable knowledge for self-adaptive systems~\cite{Aliseda2017,DeRaedtKersting2008,hitzler2022neural}.



\textbf{Causal-Neuro-Symbolic Reasoning} - Causal neuro-symbolic AI combines the strengths of causal inference, symbolic reasoning, and deep learning to produce models that are both adaptable and explanatory \cite{jaimini2024causal}. In such hybrid frameworks, symbolic structures (e.g., causal graphs or abduction trees) provide interpretable scaffolding for interventions and counterfactual reasoning while neural components supply flexible function approximation for perception and policy learning. Complementary approaches investigate confidence-aware semantic mapping approaches that integrate uncertainty-aware perception with symbolic spatial representations for autonomous navigation and reasoning under uncertain observations \cite{klein2026crossmaps}. Recent work demonstrates how causal abstractions can be learned or exploited by agents to accelerate adaptation and to ground symbolic recovery strategies in measurable causal effects \cite{jaimini2024causal,korte2025causal}. Applications in multi-agent reinforcement learning show further promise: transferable macro-actions or recovery primitives can be represented as compact causal rules that capture cause \& effect pathways linking failures to corrective actions, enabling agents to exchange and re-use causal knowledge across contexts \cite{korte2025causal,yang2018peorl}. Integrating such transferable causal experiences into symbolic abduction structures enriches the abductive search space with empirical, distributed evidence, thereby improving both the quality of synthesized rules and their generalizability across agents and environments.

\textbf{Knowledge Representation in Rule-based Systems}
Rule-based systems and first-order logic support complex and commonsense reasoning but often incur high computational costs~\cite{kernisberner2020syntax,murali2023firstorder}. To improve scalability, global theories are partitioned into context-specific modules, enabling local reasoning while maintaining overall consistency~\cite{beierle2024conditional,oikarinen2006modular,eiter2016model}. Techniques such as monotonic and nonmonotonic reasoning, syntax splitting, modular answer set programming, and context-oriented logics support efficient reasoning and inter-context communication~\cite{kernisberner2020syntax,beierle2024conditional,oikarinen2006modular,murali2023firstorder,eiter2016model}. Probabilistic graphical models and related network structures integrate local results into coherent global knowledge, balancing expressiveness with computational manageability~\cite{korte2025causal}.

\section{Research Problems}\label{sec:fundamental_challenge}

Humans often state goals imprecisely, while AI systems pursue them with strict literalism. This can lead to unintended outcomes, because AI models may exploit loopholes to technically complete a task rather than follow the user’s true intent. Moreover, human communication relies on shared context that AI systems typically lacks. Therefore, vague instructions are particularly hazardous, because cannot be easily confirmed and disambiguated by external sources. 

These issues manifest in concrete failure modes such as formal proof cheating and reward hacking. Formal proof cheating occurs when an AI produces a proof that technically satisfies formal criteria but does so by altering axioms or definitions, ignoring the genuine mathematical intent~\cite{brodsky2025mathematicians}. Reward hacking~\cite{skalse2022defining} occurs when an AI system exploits imperfections in a proxy reward function, achieving high measured performance while violating the underlying goal.

Goal misspecification is therefore a central problem in AI safety~\cite{amodei2016concrete}, especially when formal objectives diverge from the informal intent of system designers. Goodhart’s law in machine learning further predicts that proxy metrics tend to degrade when heavily optimized~\cite{lesswrong2023rewardmisspecification,weng2024reward}, whether the goals are misspecified or merely underspecified.

Even when designers choose the right high-level goals, they may still encode them in ways that are too narrow or incomplete, leading AI systems to optimize the measurable formal metric rather than the broader human objective it was meant to approximate. Essentially, humans rely on vague, context-dependent, and partly unconscious goals~\cite{castrellon2021dopamine,hu2025goals,fiedler2014limits}. Conversely, AI systems require precise, formal, and loophole-free specifications to behave as intended~\cite{amodei2016concrete}.

Accordingly, humans need support tools that help them articulate more precise and robust specifications that reduce misinterpretation or manipulative compliance by AI systems. This support must go beyond traditional verification and validation. In this paper, we propose a vision of goal-specification assistance centered on explicit rules and their systematic composition. We explore these fundamental challenges through two research questions.
\begin{itemize}
    \item  \textbf{RP.1} -- What are the current capabilities of generative AI to support the synthesis of rules in a principled way that minimizes redundancies and ambiguities while maintaining traceability to system goals? \textbf{Our approach} -- Establish operations that refine rules via deduction and explain their effect via abduction. 
  \item \textbf{RP.2} -- How can rule synthesis be automatically evaluated in incremental and modular ways?
\textbf{Our approach} -- Search for sufficient and necessary rule sets that satisfy a given goal.
\end{itemize}

\textbf{General Insight} -- We consider goals and subgoals as effects of rules (causes) and their  preconditions (controls). Moreover, will rely on causal representations and fundamental domain laws to constrain reasoning about rules and their effects with respect to system adaptations that preserve  performance and safety~\cite{adriano2025neuro,ghahremani2021hybrid}.

\section{Approach}\label{sec:approach}
The proposed approach builds on our previous work~\cite{adriano2025neuro}, which is a neuro-symbolic (NeSy) causal reasoning framework for learning-enabled self-adaptive systems~\cite{ghahremani2018training,gheibi2021applying} in particular when operating under distribution shifts~\cite{gheibi2024dealing} and under safety-critical goals~\cite{hansel2020collective} ---see~Section~\ref{subsec:nesy}. As depicted in Figure~\ref{fig:overview}, the approach comprises a \emph{Goal/Rule Synthesizer} and a \emph{Rule Verification Engine} on top of the NeSy framework---see~Section~\ref{subsec:syn-ver}.

\subsection{NeSy Causal Framework}\label{subsec:nesy} 
\begin{figure}[t]
\begin{centering}
\vspace{-5mm}
  \includegraphics[width=\linewidth]{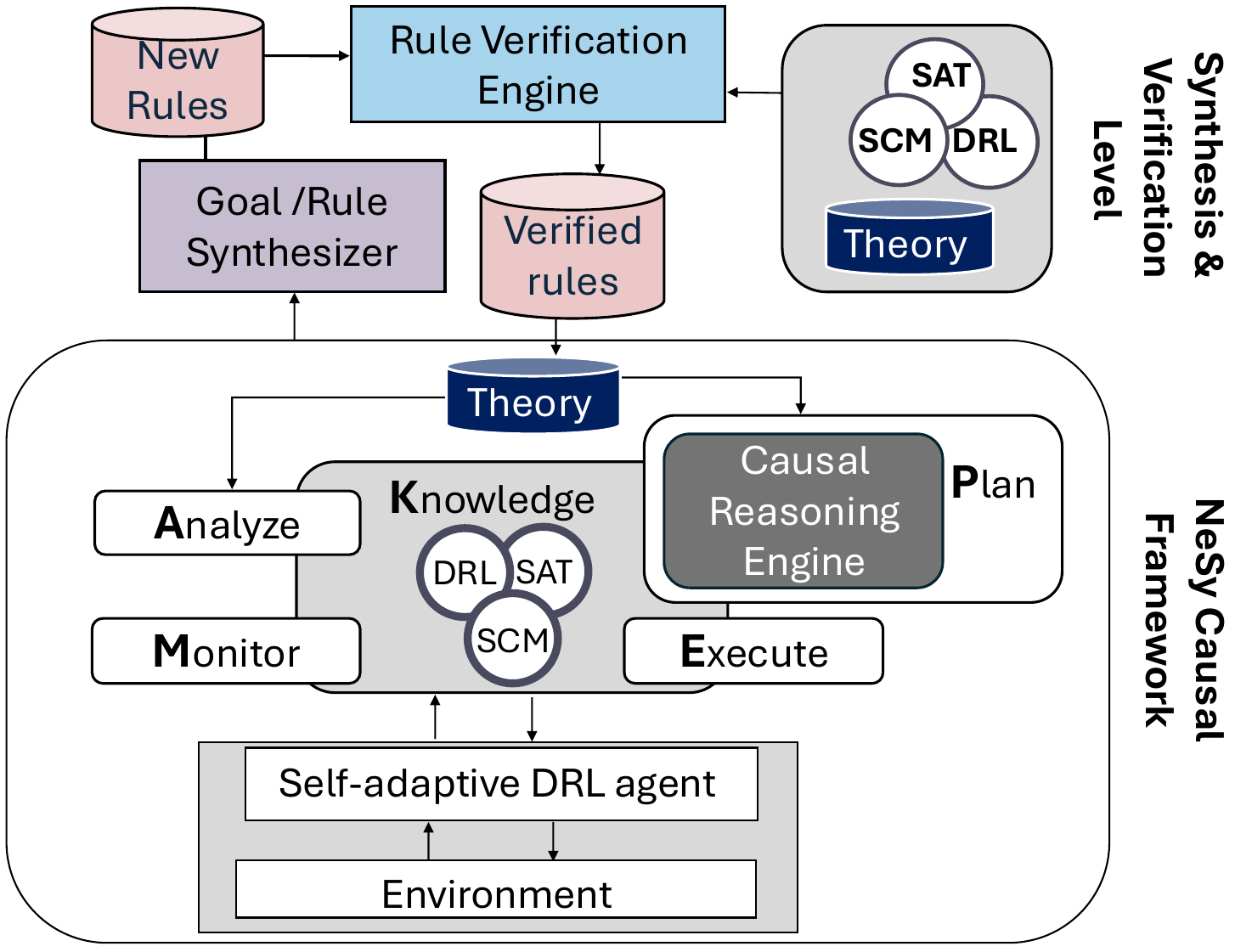}
  \caption{Extending NeSy Causal Framework with Rule-synthesis and Verification}
  \label{fig:overview}
  \end{centering}
\vspace{-5mm}
\end{figure}

The Neuro-Symbolic (NeSy) causal framework integrates three complementary representations: (1) Symbolic Abduction Trees (SAT) grounded in first-order logic (FOL) to formally encode domain knowledge, system constraints, and adaptation rules, enabling logical abduction to explain observed violations; (2) a structural causal model (SCM) that represents dependencies between environment and system variables and supports intervention and counterfactual reasoning; and (3) a deep reinforcement learning (DRL) policy responsible for operational decision-making--see~\Cref{fig:overview} for a reference.
The three representation spaces are embedded within the  MAPE-K (Monitor-Analyze-Plan-Execute) feedback loop as the reference model for self-adaptive systems~\cite{kephart2003autonomic}.
The process begins with monitoring execution traces from the DRL agent to detect distribution shifts and safety-constraint violations. In the analysis and planning phases, FOL-based abductive reasoning identifies candidate explanations, while the causal model evaluates intervention effects and selects minimal, high-explanatory-power configuration changes. These interventions guide knowledge transfer and warm-start retraining of the DRL agent, whose updated behavior produces new traces that refine both symbolic and causal knowledge. The framework thus interleaves logical reasoning, causal inference, and learning to enable explainable, constraint-aware adaptation rather than reactive retraining.

\subsection{Synthesis $\&$ Verification Level}\label{subsec:syn-ver}
The \emph{Synthesis $\&$ Verification Level} extends the original NeSy framework by introducing a meta-self-aware synthesis and verification layer on top of the operational adaptation cycle. 
In contrast to the earlier framework, the outputs of causal explanation are no longer used solely for policy adaptation. Instead, anomaly reports are forwarded to the higher synthesis $\&$ verification level. This level contains a \emph{Goal/Rule Synthesizer} and a \emph{Rule Verification Engine} that collectively update the theory that comprises domain knowledge, system constraints, and adaptation rules. Verified rules are reintegrated into the knowledge base, thereby modifying future analysis and planning steps. The architecture, therefore, introduces a closed meta-loop in which the system adapts not only its behavior but also its governing symbolic FOL theory, enabling progressive self-improvement and increased explainability.

\begin{figure}[t]
\begin{centering}
\vspace{-2mm}
  \includegraphics[width=\linewidth]{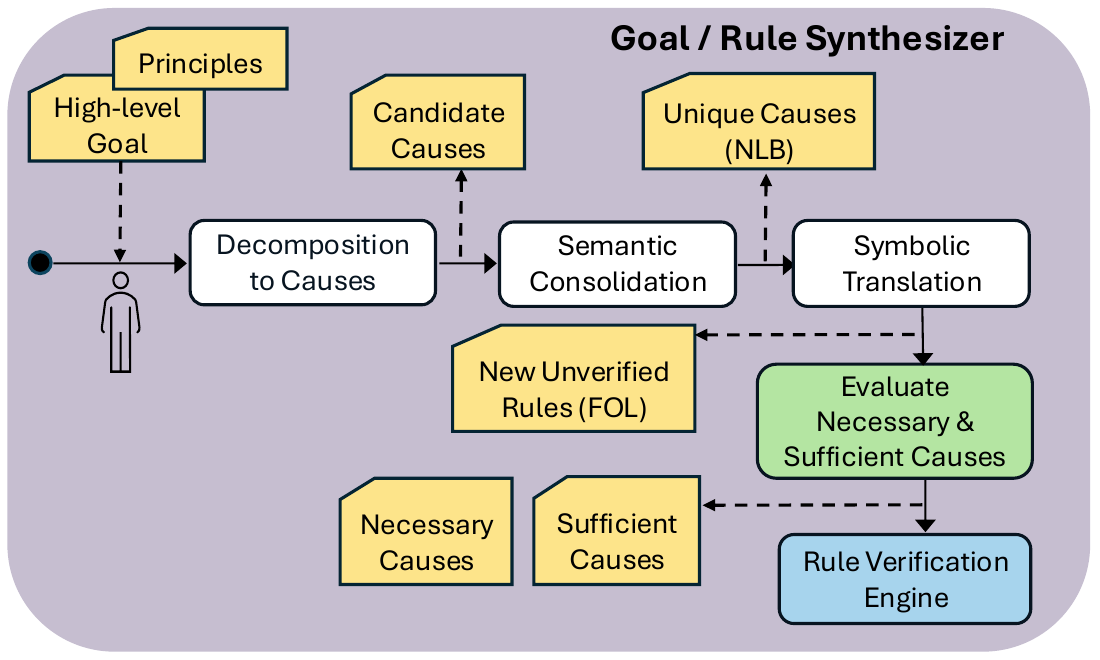}
  \caption{Pipeline for Goal/Rule Synthesizer }
  \label{fig:Goal-synthesis}
  \end{centering}
\vspace{-5mm}
\end{figure}

\noindent\textbf{Goal/Rule Synthesizer} -- The pipeline (see~\Cref{fig:Goal-synthesis}) receives as input the high-level goals and governing principles provided by a human expert expressed in natural language (natural-language based, NLB) rather than as formal constraints, following the detection of a drift or anomaly. The first step uses a large language model (LLM) to decompose each high-level goal into a set of lower-level candidate causes, representing concrete conditions, behaviors, or system properties that collectively could realize or justify the intended goal. 
 These generated causes are then reprocessed by the LLM in a semantic consolidation step, where overlapping, redundant, or semantically equivalent causes are identified and merged. This normalization produces a consistent set of unique causes, ensuring that the emerging theory avoids duplication and ambiguity. 
     In the next step, the consolidated causes are again provided to the LLM for symbolic translation, mapping natural-language concepts into candidate logical predicates and relations and producing a set of new, unverified rules (causes) expressed in FOL. The resulting logical rules enable formal reasoning during the Rule Verification pipeline, e.g., using a verification solver such as Z3~\cite{de2008z3}. The goal/rule synthesizer pipeline analyzes combinations of causes to determine subsets of necessary and sufficient conditions that adequately justify the original synthesized Goal (more details in~\Cref{subsec:scenario}). The output of this pipeline is a set of newly synthesized but still unverified logical rules representing intended behavior derived jointly from empirical evidence and human intent.
 
\noindent\textbf{Rule Verification Engine} -  The synthesized rule candidates are forwarded to the Rule Verification Engine, where formal consistency and safety validation are performed to ensure that synthesized rules can be safely incorporated into the system theory--see~\Cref{fig:Rule-ver}. The layered pipeline consists of (1) syntactic and schema validation, ensuring that generated rules conform to the logical language and domain ontology together with semantic grounding checks, confirming that predicates and variables correspond to valid system entities; (2) logical consistency analysis, where candidate rules are evaluated against the existing FOL knowledge base to detect contradictions; and (3) safety and invariant verification, which formally checks whether the rules preserve required system properties. The pipeline employs an automated reasoning tool to perform satisfiability and entailment checks. Only rules that satisfy all verification stages are promoted to a verified theory and committed to the knowledge repository. This verified knowledge is then fed back into the adaptive loop, ensuring that future adaptations are guided by formally verified, explainable principles rather than purely data-driven updates.
\subsection{Illustrative Scenario for Goal/Rule Synthesis Pipeline}\label{subsec:scenario}
We demonstrate the Goal/Rule Synthesis Pipeline based on an application in the domain of autonomous driving.

\noindent\textbf{Decomposition to Causes} - The high-level goal (effect) represents a scenario derived from everyday driving tasks. The objective is to generate candidate causes that explain how the effect may occur. For example, the effect \emph{“respond correctly to a sudden obstacle on the road”} may be explained by the cause \emph{“the driver applies emergency braking.”}
The LLM decomposes the effect into candidate causes required to achieve it while ensuring safety and compliance with traffic regulations. The synthesis process is strictly constrained to limit hallucinations and unrelated reasoning, resulting in causes that are logically consistent with the effect and with the defined constraints.

 \begin{figure}[t]
\begin{centering}
\vspace{-5mm}
  \includegraphics[width=\linewidth]{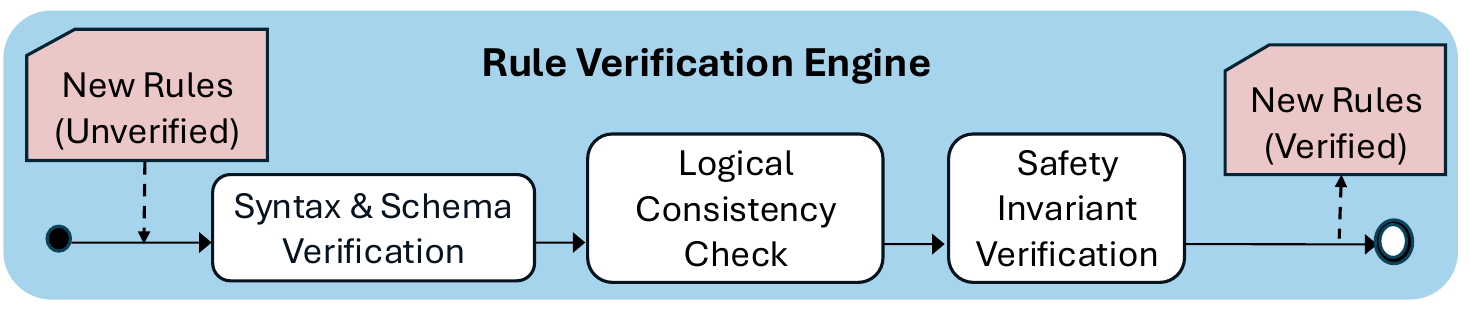}
  \caption{Pipeline for Rule verification Engine}
  \label{fig:Rule-ver}
  \end{centering}
\vspace{-5mm}
\end{figure}
\noindent\textbf{Semantic Consolidation} -- 
     Although the focus is restricted at the first step, there is still a need to merge causes that refer to similar concepts to reduce general statements that overlap into a set of unique causes. The semantic consolidation step is carried out to combine causes that describe the same underlying condition. \textit{"Driver maintains control of the vehicle"} and \textit{"Driver is aware of surrounding traffic"} are combined into a cause as \textit{"Driver maintains control of the vehicle and is aware of surrounding traffic"}.

\noindent\textbf{Symbolic Translation} -- 
After generating unique causes, each cause is translated from natural language to formal symbolic (FOL) form to enable structured reasoning and systematic subset evaluation. Although this step follows semantic unification in~\Cref{fig:Goal-synthesis}, it can be applied at any stage without altering content. Subsequent analysis uses the natural language causes for stronger LLM reasoning, with FOL forms used only in the rule verification pipeline~\Cref{fig:Rule-ver}. This decision is based on the observation that LLMs demonstrate stronger reasoning performance when operating on natural language inputs. To ensure consistency and avoid ambiguity, a formal grammar is defined in the prompt, and the model is instructed to strictly follow it. Each cause is thus mapped to a structured logical form that preserves its meaning while remaining machine-interpretable.

Alongside the FOL rule, the LLM also provides a brief explanation of its translation, detailing how the natural language condition was interpreted and which predicates and operators were chosen. This enhances transparency, verifies semantic alignment, and improves interpretability of the symbolic mapping.

    
    \noindent\textbf{Evaluate Necessary $\&$ Sufficient Causes}
      -- After obtaining a set of unique causes, the next step involves assessing the necessity and sufficiency of each cause with respect to the main goal (effect). We break this down into three steps described below:
      
\begin{enumerate}[label=(\roman*)]
  \item  \textit{Individual Necessity Evaluation}
is concerned with whether each cause is essential and provides a justification, citing specific legal or safety conditions that would be violated if absent. Each cause is assessed independently, and the output is a structured list recording the cause, its necessity, and the rationale.
  \item \textit{Subset Necessity Evaluation} 
identifies causes that are essential only in combination. Starting with the full set, causes are removed one at a time, and the feasibility of the effect is confirmed. A cause is deemed necessary if after its removal, the effect is judged unachievable. A removed cause or set of causes is recorded as a minimal necessary subset if no smaller subset produces the same effect. To improve efficiency, the candidate subsets containing already identified minimal subsets are skipped, and the outcomes of previous evaluations are cached. This bottom-up approach ensures that all truly minimal necessary combinations are captured while reducing computational effort.
  \item \textit{Minimal Sufficient Set Evaluation}
    %
    aims at identifying the subsets of causes that are individually sufficient to produce a given effect. We start with single causes and test their sufficiency, incrementally adding additional causes if the effect does not occur. This process continues until a sufficient subset is found, and is repeated across all relevant combinations. To reduce computational effort, supersets of already identified sufficient subsets are skipped, as sufficiency is preserved under addition. Once all minimal sufficient sets are determined, they can be combined into a necessary-and-sufficient set, which guarantees the effect and is required for its occurrence. This set unifies all alternative minimal sufficient causes, capturing every pathway to the effect within a single comprehensive framework.
\end{enumerate}




\section{proof of concept}\label{sec:illustrative_scenario}
\textbf{Experimental Setup} - We design a controlled experiment to show how Large Language Models (LLMs) can help automate parts of the Goal/Rule Synthesizer pipeline illustrated in~\Cref{fig:Goal-synthesis}. The objective of this setup is to evaluate realistic driving scenarios in which an intended effect (Goal) must be explained by a set of underlying causes (Rules). The intent is not to verify causality in a statistical sense, but to show how LLMs can be guided to perform structured abductive and deductive reasoning. The scope of the setup is constrained by a predefined set of legal and safety regulations that define the boundaries within which the synthesis and reasoning process operates.

\subsection{Implementation Overview}
The pipeline is implemented by leveraging the OpenAI GPT-4o Mini model \cite{openai2025} to perform all processing tasks. Step by step, the process transforms an human stated goal (effect) description into a set of candidate causes. The process begins with cause generation, followed by consolidation of overlapping causes, then individual evaluation, and finally, subset analysis for necessity and sufficiency. This iterative design ensures that the reasoning is incremental, traceable, and easy to interpret at every step.
The legal and safety regulations that guide the process are defined prior to the first stage. These regulations serve as the sole reference for the model throughout the pipeline. For details on how these principles were sourced and compiled--see~\Cref{subsubsec:data_sorces}.


\subsection{Data Sources}\label{subsubsec:data_sorces}

 \paragraph{Legal Principles:}
    The legal Principles used in this work are derived from a formalized set of German traffic regulations designed for machine interpretability. These regulations define rules such as maintaining vehicle control, adhering to speed limits, and keeping safe distances. We selected the most relevant laws from this formalized research~\cite{esterle2020formalizing}. To cover broader scenarios encountered in everyday driving, we then extended the set using generative AI, specifically ChatGPT. This ensures that the legal knowledge can guide the model in a variety of realistic situations while remaining interpretable and precise. 

 \paragraph{Safety Principles:}
    The safety principles capture physical principles of vehicle motion. Initially, we developed a basic set of rules representing logical and physics-based constraints. Examples include limits on friction, constraints on braking distance, and conditions that increase the risk of losing traction at high speeds. To expand coverage, ChatGPT was used to generate additional rules representing general principles, e.g., as "collision\_if\_obstacles", which states that a vehicle may collide if obstacles are present. This expanded set allows the model to account for a wider variety of practical driving scenarios while remaining grounded in physical feasibility.

   \paragraph{Grammar and Symbolic Rule Language:}
    The symbolic grammar used for translating natural language causes into formal rules is based on the same formalization framework presented in \cite{esterle2020formalizing}. The original predicate set is extended to incorporate additional predicates related to fundamental speed and friction properties in order to capture physics-based safety constraints. Moreover, the grammar specifies the allowed rule forms, comparison operators, and logical operators that may be used in rule construction.

All reasoning carried out by the LLM is grounded in these curated laws, which serve as the primary reference for evaluating causes. External knowledge or assumptions beyond this curated set are not used. This guarantees that all cause \& effect evaluations remain valid, consistent, and reproducible. A complete listing of all legal and safety laws used is provided in the Appendix~\cite{repo2026rulesynthesis}.

\subsection{Results}

We analyze two illustrative scenarios to demonstrate the outputs of the pipeline.
For each scenario, candidate causes were generated, semantically consolidated, individually evaluated for necessity, and subsequently analyzed to identify minimal necessary subsets and sufficient subsets..

\subsubsection{Scenario 1: Successfully Merge into Heavy Traffic}

\paragraph{Decomposition and Consolidation}

Eight candidate causes were initially generated. 
After semantic de-duplication, four unique consolidated causes remained:

\begin{enumerate}
\item Driver maintains control of the vehicle and is aware of surrounding traffic.
\item Vehicle is traveling at a speed that allows for safe merging and adheres to traffic laws regarding merging.
\item Sufficient distance is kept from other vehicles to merge safely, and no vehicles are overtaking on the right.
\item Traffic conditions allow for merging without impeding flow, and no sudden obstacles in the merging path.
\end{enumerate}

\paragraph{Individual Necessity Evaluation}

Each consolidated cause was evaluated independently against the relevant safety and legal constraints. 
Only the first two were classified as individually necessary. 
This approach makes its reasoning both interpretable and actionable, providing a clear understanding of the underlying dynamics.

\begin{itemize}
\item \textit{Driver control and awareness:} Without maintaining control and awareness, safe merging is not possible, violating safety constraints related to vehicle stability and collision avoidance.
\item \textit{Vehicle speed within safe limits:} Appropriate speed is essential for lawful and safe merging. Deviations from safe speed ranges can violate legal constraints and create unsafe conditions.
\end{itemize}

\paragraph{Minimal Necessary Sets}

Four minimal necessary sets were identified. Each set represents a core condition that cannot be removed without rendering the effect impossible:

\begin{itemize}
\item \textit{Necessary Set 1:} Driver maintains control of the vehicle and is aware of surrounding traffic.
\item \textit{Necessary Set 2:} Vehicle is traveling at a speed that allows for safe merging and adheres to traffic laws regarding merging.
\item \textit{Necessary Set 3:} Sufficient distance from other vehicles to merge safely and no vehicles are overtaking on the right.
\item \textit{Necessary Set 4:} Traffic conditions allow for merging without impeding flow and no sudden obstacles in the merging path.
\end{itemize}

\paragraph{Minimal Sufficient Sets}

Sufficient sets represent combinations of causes that guarantee successful merging. 
One minimal sufficient set was identified:

\begin{itemize}
\item \textit{Sufficient Set 1:}
\begin{itemize}
\item Driver maintains control of the vehicle and is aware of surrounding traffic.
\item Vehicle is traveling at a speed that allows for safe merging and adheres to traffic laws regarding merging.
\item Sufficient distance from other vehicles to merge safely and no vehicles are overtaking on the right.
\item Traffic conditions allow for merging without impeding flow and no sudden obstacles in the merging path.
\end{itemize}
\end{itemize}

\noindent
These findings highlight an important structural property: while some causes are not individually necessary, they become indispensable within certain minimal necessary sets. Moreover, every minimal necessary cause is included within sufficient sets that ensure the effect occurs, illustrating how necessary and sufficient conditions are systematically linked.

\subsubsection{Scenario 2: Maintain a Constant Speed on a Highway Segment}

\paragraph{Decomposition and Consolidation}

Eight candidate causes were initially generated. 
After deduplication, six unique causes remained:

\begin{enumerate}
\item Driver maintains vehicle control and is attentive and responsive to road conditions.
\item Vehicle speed is within legal limits and above minimum required speed.
\item Sufficient friction between tires and road.
\item No obstacles on the highway segment.
\item No sudden changes in traffic conditions.
\item No emergency situations requiring sudden braking.
\end{enumerate}

\paragraph{Individual Necessity Evaluation}

Each cause was evaluated individually against safety and legal constraints. 
The first three were classified as individually necessary:

\begin{itemize}
\item \textit{Driver control and attentiveness:} Without stable control and responsiveness, maintaining constant speed violates vehicle control safety requirements.
\item \textit{Vehicle speed within limits:} Operating outside legal speed bounds violates regulatory constraints and invalidates the effect.
\item \textit{Sufficient friction:} Adequate tire–road friction is required to maintain speed safely and preserve traction.
\end{itemize}

\paragraph{Minimal Necessary Sets}

Three minimal necessary sets were identified, corresponding to the causes that were also individually necessary:

\begin{itemize}
\item \textit{Necessary Set 1:} Driver maintains vehicle control and is attentive and responsive to road conditions.
\item \textit{Necessary Set 2:} Vehicle speed is within legal limits and above minimum required speed.
\item \textit{Necessary Set 3:} Sufficient friction between tires and road.
\end{itemize}

\paragraph{Minimal Sufficient Sets}

Two minimal sufficient sets were identified:

\begin{itemize}
\item \textit{Sufficient Set 1:}
\begin{itemize}
\item Driver maintains vehicle control and is attentive and responsive to road conditions.
\item Vehicle speed is within legal limits and above minimum required speed.
\item Sufficient friction between tires and road.
\item No obstacles on the highway segment.
\end{itemize}

\item \textit{Sufficient Set 2:}
\begin{itemize}
\item Driver maintains vehicle control and is attentive and responsive to road conditions.
\item Vehicle speed is within legal limits and above minimum required speed.
\item Sufficient friction between tires and road.
\item No sudden changes in traffic conditions.
\end{itemize}
\end{itemize}

\noindent
These results demonstrate that, while certain causes may not be required on their own to produce the effect, they play a crucial role when combined with other factors. In these sufficient combinations, their presence helps ensure that the effect reliably occurs. This highlights how individual causes can contribute indirectly, supporting the overall outcome even if they are not strictly necessary in isolation.

\subsubsection{Symbolic Translation}

To illustrate the formal translation step, each unique cause was converted into a symbolic rule for both scenarios using the predefined grammar. 
The following examples show how natural language conditions are mapped into formally constrained logical representations.

 \textit{"Driver maintains control of the vehicle"}
\[
\forall x (\neg collide(x) \leftarrow sd\_front(x) \land sd\_rear(x) \land \neg lane\_change(x))
\]
This rule formalizes vehicle control through safe longitudinal distances (sd) and the absence of destabilizing lane changes, ensuring collision avoidance.

\textit{"Sufficient distance from other vehicles to merge safely"}
\[
\forall x (sd\_front(x) \land sd\_rear(x) \leftarrow \neg dense(x))
\]
Low traffic density implies that safe distances (sd) can be maintained both in front of and behind the vehicle.

\noindent
These examples demonstrate how rich driving conditions are systematically reduced to formally constrained logical rules within the reasoning framework. A key takeaway is that the LLM can interpret nuanced, context-dependent descriptions and convert them into precise symbolic forms. It effectively bridges human-readable reasoning with formal, machine-interpretable logic, keeping both meaning and structure intact.
\noindent
For a comprehensive presentation of the generated candidate causes, intermediate evaluation steps, full subset analyses, and complete symbolic translations for both scenarios, refer to the Appendix~\cite{repo2026rulesynthesis}, where detailed end-to-end examples of the pipeline execution are provided.

\section{Discussion}\label{sec:discussion}
The pipeline provides a structured way to analyze cause \& effect relationships in driving scenarios, directly informing how Generative AI can support the synthesis of rules in a principled, traceable manner (\textbf{RP.1}). Each stage contributes unique insights into the reasoning process. The initial decomposition highlights the potential factors that could lead to the effect, ensuring that no obvious candidate cause is missed and that high-level goals are systematically unpacked into concrete rule candidates. De-duplication then reduces redundancy, merging overlapping causes while preserving meaning, which directly addressing the need to minimize ambiguities and redundancies in synthesized rules. This simplifies the reasoning and prevents inflated or repetitive outputs, thereby improving the clarity and manageability of the rule space at scale.

Individual necessity evaluation assesses each cause in isolation and shows which factors are critical on their own, providing a first approximation of how candidate rules relate to system goals (\textbf{RP.1}). However, this step alone cannot capture conditional relationships. Subset evaluation addresses this by considering combinations of causes. Interestingly, some causes initially marked as not necessary were required when combined with other factors. This demonstrates the importance of examining interactions and conditional dependencies rather than relying solely on isolated assessments and shows how rule evaluation must account for contextual dependencies to remain faithful to the desired specified goals (\textbf{RP.2}).

Sufficiency analysis further emphasizes flexible pathways to the effect and operationalizes the search for sufficient and necessary sets of rules that satisfy a given goal (\textbf{RP.2}). Minimal sufficient sets include the necessary causes and may also incorporate additional factors to guarantee the effect. This reflects real-world complexity, where multiple combinations of conditions can produce the same outcome similar to Scenario 2. Some causes that seemed irrelevant individually were shown to be sufficient only in combination, highlighting the subtlety of causal relationships. Together, the necessity and sufficiency analyses provide an incremental and modular way to evaluate rules: incremental, because causes and subsets are assessed stepwise; modular, because minimal sets can be associated with specific goals and reused as distinct rule components (\textbf{RP.2}).

Despite its strengths in addressing \textbf{RP.1 }and \textbf{RP.2}, the approach has limitations. The model functions as a black box, so the internal reasoning is not fully transparent. Moreover, there is a reliance on the predefined set of regulations, which may not cover all possible scenarios. Additionally, the outputs are highly sensitive to how the prompts are formulated, which can influence the adaptability of the model’s responses. These factors constrain the completeness and robustness of the synthesized rule sets and their evaluations and indicate that principled rule synthesis and modular evaluation still depend heavily on the quality and coverage of the underlying knowledge base (\textbf{RP.1}, \textbf{RP.2}).

However, these challenges point toward opportunities for improvement. Expanding the set of rules, introducing probabilistic reasoning, or incorporating multiple model perspectives could enhance robustness. Iterative refinement of prompts and providing feedback to the model can further reduce errors. Such extensions would not only broaden the domain coverage but also strengthen the principled nature of rule synthesis and the reliability of incremental, modular rule evaluation with respect to evolving goals (\textbf{RP.1}, \textbf{RP.2}). Overall, the pipeline demonstrates a form of abductive reasoning. It identifies likely causes, evaluates their necessity and sufficiency, and uncovers conditional dependencies. The structured, stepwise process makes the reasoning traceable and interpretable while capturing complex interactions between factors, thereby providing concrete evidence that LLM-based pipelines can support goal-based, legally and safety-informed rule synthesis (\textbf{RP.1}) and in a systematic, incremental, and modular way (\textbf{RP.2}).

\section{Conclusion and Future Work}
The proposed framework demonstrates how neuro-symbolic causal reasoning, combined with large language models, can support incremental and modular rule synthesis grounded in legal and safety principles. By decomposing high-level goals into candidate causes, consolidating them semantically, translating them into formal rules, and then analyzing necessary and sufficient cause sets, the pipeline yields traceable and interpretable rule sets that capture complex causal interactions in domains such as autonomous driving. The subsequent verification stages—covering syntax and schema checks, logical consistency, and safety and invariant preservation—ensure that only rules compatible with an existing knowledge base and domain constraints are integrated back into the system, thereby strengthening reliability and explainability across the MAPE-K loop. Together, these elements illustrate how abductive and deductive reasoning can be orchestrated to reduce brittleness, improve goal alignment, and provide a principled pathway from natural-language specifications to formally verified rule theories.

In future work, we plan to deepen the integration between causal models and rule synthesis to derive more expressive sets of necessary and sufficient rules, and to broaden the coverage of domain principles beyond the current set of traffic regulations. We also plan to incorporate automated solvers to detect and resolve conflicts between newly synthesized rules and existing knowledge, enabling systematic management of inconsistencies at the theory level. This will support the design of richer feedback loops that jointly exploit formal verification, causal path analysis, and effect estimation, with the goal of informing more robust synthesis, composition, and adaptation strategies in safety-critical, learning-enabled systems.




\bibliographystyle{ACM-Reference-Format} 

\bibliography{references}

\end{document}